\documentstyle[fleqn,12pt]{article}

\normalsize

\begin{document}

\section*{Comment on ``An Evolutionary Picture for Quantum Physics'' 
by Rudolf Haag}

\centerline{\bf Walter Fusseder}

\centerline{Sektion Physik der Universit\"{a}t M\"{u}nchen,} 

\centerline{Theresienstrasse 37,
80333 M\"{u}nchen, Germany} 

\centerline{email: fusseder@gsm.sue.physik.uni--muenchen.de}

\begin{abstract}
Widespread unjustified views on the role of
the observer, the individuality of quantum processes, 
the relation between decoherence and irreversibility, Bell's quest for 
`beables', the direction of time, and the concept of 
experience are revealed, whereby a better understanding is achieved of what 
the evolutionary picture is all about and what is regarded as missing in
already existing realistic pictures.
\end{abstract}

{\it 1. Why was \cite{Haa 96} published in Communications in Mathematical 
Physics?\/}
The idea was not to formulate a new theory, nor even get to grips with
the inherent difficulties of the proposed research 
program so that a theory could be formed.
An unquestionable benefit of \cite{Haa 96} is the brave confrontation with
widespread positivistic doctrines, especially to put forward the
convictions which shall lead to a new picture of quantum world---an 
evolutionary picture (EP).
It is the purpose of this comment to reveal the prejudices hidden in the EP
and to compare with already existing realistic pictures, in order to 
understand:
what is the EP about?
The clear answer given in \cite{Haa 96} (all further quotations are from 
source \cite{Haa 96}) ``an evolving pattern of events with causal links
connecting them'' is not taken seriously---oddly enough---because 
``idealizations cannot be avoided''.
Haag neglects giving definitions of basic concepts;
he even argues that this would only be possible in ``asymptotic 
situations'', whereby the EP keeps up the vagueness of orthodox quantum 
theory (OQ).
Also the description of OQ is imprecise---even in sensitive fields.
It is not clear which of the various Copenhagen interpretations is on the 
agenda---a decisive matter, because Haag's main motivation is ``the desire to 
separate the laws of Quantum Physics from the presence of an observer''.

{\it 2. The EP and Bohr's epistemological analysis.\/}
The central feature of the EP is an event: the ``transition of a possibility
to a fact''.
From Bohr's epistemological analysis of quantum processes  
Haag adopts ``indivisibility''---an apparently reasonable step.
No attempt is made to draw a picture that would give a
``subdivision of such processes...an objective meaning''.
Instead an event generalizes the notion of measurement result because it
``is considered as a fact independent of an observer''.
Haag's description of ``the narrow view'' of OQ is inappropriate, and
lots of the unjustified features of the EP arise from these 
misunderstandings.

One can say, contrary to the usual concentration on the objective world, 
classical physics (CP) is the projection of the objective and timeless
properties of reason (like logic and mathematics) on the outside world.
As forms of intuition ({\it Anschauungsformen\/}) and  conceptions of
understanding ({\it Verstandesbegriffe\/}), basic notions like space, 
time, and causality are properties of cognitive faculties.
The subject that constitutes CP is not part of nature
and does not depend on the flow of time. 
It is faced with nature:
thinking and decisions of free will are independent of nature.
If CP is taken for granted, experience is the cognition of 
objects by perception, i.e.~the classical interpretation of an
observation.
Bohr of course analyzed atomic processes according to this concept of 
experience.
As a result he challenged epistemology:
{\it How can experiments with atomic systems constitute experience?\/}
Strictly speaking, it is impossible to 
interpret these experiments with classical notions, i.e.~causally in space and 
time;  
they give only complementary fragments of experience, which can be translated 
into the symbolism of OQ, 
but classical experience cannot be composed of these fragments.
In this context the notion of individuality should indicate that within a 
measurement intermediary stages cannot be interpreted classically.
Further, OQ teaches that for processes with nonnegligible action
it is not possible to design a real outside world: 
``There is no quantum world'' (Bohr).
This `teacher', however, is not a positivistic prejudice about unobserved 
quantities as Heisenberg's early writings announced effusively,
but the notion of 
experience that presupposes realistic, namely classical, theories.

Von Neumann agreed in no way with such dogmatism about CP; 
his principal concern was to abandon causality.
He introduced randomness according to the expression `disturbing of 
phenomena by observation', which was later rejected by Bohr as misleading.
Strictly speaking, von Neumann projected the relation between nature
and subject on every {\it single\/} observation.
On the pretext of the principle of psycho--physical parallelism he justified 
the collapse postulate as an image of the indescribable (but nevertheless 
incompatible with the assumed principle) interaction between subject and object.
(Its concrete formulation was abstracted from the Compton--Simons 
experiment.)

Bohr's interpretation has no problem with unobserved quantities if they 
are classical.
If the action is nonnegligible, independent reality is denied.
Different to von Neumann's quantum theory:
unobserved systems evolve according to Schr\"odinger's equation.
Observations go beyond physics.
As usual now, Bohr has to take the rap for representing an 
observer--overloaded quantum theory. 
But something similar to von Neumann's position is expressed:
``...if we want to understand the word observer in a wider sense we must endow 
him at least with the faculties of consciousness, intelligence in planning and 
free will in execution.''
Bohr stressed that it is impossible to separate phenomena and agencies of 
observation.
Consciousness is in this connection irrelevant. 
Only von Neumann's projection introduced it.
As a result they had different views on the outside world.
And, loosely speaking, Bohr's `theory' is about the wave function $\psi$ 
(not as a physical quantity) and the classical world, while von Neumann's 
`theory' is about $\psi$ and the consciousness. 
All these things are poorly expressed in \cite{Haa 96}.
Where exactly does one draw the dividing line between observer and subject?
What causes the quest for realism:
the prominent role of measurements in OQ, or the 
agencies of observation which one does not expect in CP?

More important is Haag's inconsequence concerning ``the epistemological 
analysis of Niels Bohr''.
On the one hand, he had ``no basic disagreement'' with this analysis.
On the other hand, he throws doubt upon it, because he 
``claim[s] that there are discrete, real events on any scale''.
Surprisingly, the obvious cause of this doubt---the abandonment of the `shifty'
split which is necessary for the consistency of Bohr's `theory'---is not 
mentioned.
On the contrary, everything is done to repress the doubt---by the
shady excuse that ``an individual event is an asymptotic notion''.
To elaborate---with the help of the notion of event phenomena and agencies of 
observation can be separated:
an event (generally speaking) rearranges the division between system and 
apparatus by introducing a {\it definite split\/} between an
irreducible system---where the event occurs---and its surrounding---which
serves for amplification.
In von Neumann's language:
the collapse occurs at a definite boundary.
In Bohr's language:
on what subsystem can we gain fragmentary experience with certainty?
Although the contradiction to Bohr's analysis seems to be weak---indivisibility
is not touched---`quantum world' gets a new status: 
it is `created' by the EP.
Bohr's `theory' is about the (merely symbolic) $\psi$ investigated from 
the classical outside world. 
The EP is not simply an extension which is enabled by an 
``alternative language'', but a different theory
about $\psi$ and events occurring in the nonclassical outside world.
There is no need for a classical world to constitute the quantum world.

{\it 3. Comparison with `consistent histories' (CH).\/}
CH is a theory about $\psi$ and events:
the values of observables which are 
assigned if a certain decoherence condition is fulfilled.
There are two main differences between the EP and CH.
The first one concerns time.
CH draws a timeless picture, a history, i.e.~the values of observables at a
sequence of time, has no preferred direction.
On the contrary, in the EP time is directed.
It distinguishes by an irreversible process the (extended)
moment when a possibility 
becomes a fact and with it a flow of time (the past is fixed, the future is 
open) is supposed to be introduced.
In this sense the EP is a theory about $\psi$ and events which pop up
out of ``possibilities'' (their specification and probabilities are determined
by $\psi$).
But now the question arises (corresponding to the second difference):
what is analogous to the decoherence condition?
Bohr's insistence on classical experience is thrown into doubt. 
Only an evasive answer is given:
``the hinging of basic concepts to asymptotic situations which are
only approximately realizable''.
The division of a measurement in an event and an amplification suggests the 
following strange picture:
The surrounding defines according to which observable the possibilities are 
formed out of $\psi$ (though no interaction with the surrounding occurred).
The decision for a certain possibility is made when something like a 
decoherence condition is fulfilled.
If the event is shifted to macroscopic bodies, 
it is difficult to establish a difference between the EP and `spontaneous 
localization' (SL).
But all these are speculations about a non--existent theory, strictly speaking 
about the mentioned shady excuse.

{\it 4. Are there ``events on any scale''?\/}
It is difficult to analyze a non--existent theory, but if this claim is
maintained, the probabilities of events would differ from those calculated by
OQ.
The unaffected interpretation of Schr\"odinger's equation, namely 
``it describes
the continuous change of probabilities for possible facts'', in other words
an observable {\it has\/} a certain value and its probability is calculated by
Born's (generalized) rule, nevertheless contradicts Born's rule.
The reason is simply the theorem of Kochen and Specker, which forbids the 
assignment of probabilities to too many `observables' consistent with OQ.
(Applied to the EP, this is a statement about ``possibilities''.)
Although the probabilities are calculated in the same way, they have a 
different meaning.
In the EP probabilities are no longer {\it conditional\/} probabilities 
presupposing that a measurement is actually performed.
(For an application to CH, turn to \cite{Gol 97}.)
In order to maintain consistency with OQ, events cannot exist on 
{\it any\/} scale.
The way out of OQ, whereby it is impossible to perform
arbitrary measurements simultaneously and that their results are joint products 
of system and apparatus, cannot be ignored by a realistic interpretation
and must lead to a further constraint on the notion of event---no such
thing is mentioned in \cite{Haa 96}.

{\it 5. Decoherence and irreversibility.\/}
In order to be real, it is argued in \cite{Haa 96}, events have to be 
irreversible---a new type of irreversibility.
The explanation of irreversibility by statistical mechanics is restricted to
the macroscopic world:
``If we believe that this is the only mechanism by which irreversibility can 
arise we must conclude that the elementary processes, even if isolated, cannot
be regarded as real but needs the macroscopic amplification before we can 
talk about a fact.''
Contrary to this view, Bohmian mechanics (BM) gives an example of a 
deterministic and time--symmetric theory
which clarifies the relation between decoherence and irreversibility
and which includes even events as an asymptotic notion in its phenomenology.

To explain the occurrence of facts in OQ by decoherence (without an additional
collapse postulate) is naive and ambiguous.
BM has in this respect a more striking concept (striking, because BM has no
`measurement problem') of distinguishing between 
orthogonal wave functions and wave functions with disjoint supports.
The analysis of decohering processes 
(like 
scattering of fast particles or the formation of bubbles in a bubble chamber)
shows that those producing wave functions with (nearly) disjoint supports 
(like scattering of fast particles or the formation of bubbles in a bubble 
chamber) are practically irreversible.
(You cannot catch up with a photon that is flown away,
a macroscopic number of degrees of freedom is uncontrollable.)
Events are defined---with familiar mathematics---as the inverse images
of random variables (deterministic functions on configuration space).
They can be regarded as an asymptotic notion in the following sense:
typically the range of a random variable is a continuum, but in a certain 
limit (e.g.~if the velocity of a scattered particle or the number of 
molecules in a bubble tends to infinity) the range consists of discrete values,
which become strictly correlated to an appropriate (macroscopic) variable 
defined on the configuration space of the surrounding.
Which of these discrete values ``is selected by nature in the individual case'' 
is traced back (deterministically) to the initial configuration.

Why should Bohr be alleged to have wanted to introduce
a notion of {\it strict\/} irreversibility?
Irreversibility in the context of observation should
emphasize the gap between classical and quantum mechanics:
interference destroyed in the (necessary) amplification up to the classical 
domain cannot in practice be recovered.
Also Pauli---well--known for his judgement of fundamental
questions and never in conflict with Copenhagen---never mentioned 
irreversibility in his textbook, but ``...during each measurement an 
interaction with the measuring apparatus ensues which is in many respects 
intrinsically uncontrollable''. 
From this point of view the justification of why events should be irreversible,
namely because events are the warranted extrapolation of irreversible
measurements, is not only vague, but also disproved by BM (if the 
necessity of irreversibility is claimed).

{\it 6. Event as the ``transition of a possibility to a fact''.\/}
Events are supposed to introduce 
an irreversible element unknown to CP.
They reveal much about the role of the subject because this new type of
irreversibility is justified by a certain feature of indeterministic theories:
facts are ``related to the past'' and ``possibilities are related to 
the future''. 
Accordingly, ``statistical statements in physics must always be future
directed''.
This results from an (apparent) interlink between probability and 
experience---affected in a Laplacian world.
Furthermore, experience is understood as learning from the past for the future.
Undoubtedly it is consistent to construct physics as an empirical science
together with a future directed notion of probability and an adjusted 
temporal logic, but in view of the timeless axioms of Kolmogoroff and 
the objective forms of cognition it seems very
unlikely that this concept of physics is {\it necessary\/}.
The intention is undoubtedly to overcome the timelessness of CP 
and to throw the subject into the flow of time.
Because of this, thinking does not become a process in nature, only the 
(plausible) psychological flow of time is projected onto nature.
This projection also seems to distinguish the (extended!) present 
moment when an event occurs, and with it part of the future becomes the past.

Neither perception of events as irreversible 
nor this concept of temporal experience seem to be convincing and necessary.
The flow of time and with it directing the links between events is as 
fictitious as in CP:
time is standing still.
The evolutionary point of view
of nature differs from the classical one only concerning determinism.
Evolution as a ``growing category'' is fictitious---produced in a projection 
of our sensation of time on physics.

{\it 7. Events as `beables'.\/}
Footnote 1 in \cite{Haa 96} gives an example of the pernicious 
influence of mathematics on science:
the adoption of formulating results as statements about limits.
As Bell pointed out these limits blur fundamental and practical questions.
A classic example is a density matrix where the off--diagonal elements
weakly vanish, if the apparatus becomes arbitrarily large.
Focussing on the system, `and' has been replaced by `or'---by an innocent 
increase of the apparatus.
Bell's ``quest for `beables' which can be precisely defined under any 
circumstances'' is the demand for a clear ontology which avoids such blurs.
Analogously, in the EP the practically
uncontrollable decoherence in a measurement is converted into strict
irreversibility.
In a theory with precisely defined `beables', e.g.~BM, no such 
blunder occurs---every unavoidable shortcoming of our models can be analyzed 
on the practical level.

And if a theory is based on the notion of irreversible events, these events
are by no means an ``asymptotic notion'', but
in every special model they are sharply defined (though extended)
in space and time in irreducible subsystems.
There is no need for a limit to define events.
The more detailed the model is designed the more precise should be the 
theoretical predictions---again all differences appear on the practical level.
To regard the vagueness of OQ---the shiftiness of the split---as a
consequence of the fact that ``with increasing insistence on precision the 
subdivision of the universe must become necessarily coarser and the description
less detailed'' shows on the contrary that events are not taken seriously
as the fundamental concept of the EP.

{\it 8. Is there any connection with Whitehead's process--metaphysics?\/}
On the one hand, since Haag claims to agree with Bohr's epistemological 
analysis, especially by adopting the concept of indivisibility, 
he supposes Bohr's concept of experience.
On the other hand, the statement that ``physics transcends 
epistemology'' supports an adaptation of experience to 
physical theory
(against the widespread view among physicists and even critics of
positivism that experience is independent of any physical theory).
The (repressed) doubt about Bohr's analysis confirms this view. 
Was Bohr too cautious when he stressed the classical behaviour of the 
apparatus?
The notion of event forces one to abandon Bohr's concept of experience.
Because of that the statement ``the raw material of physics, which the 
theory is supposed to explain, consists of facts which can be documented''
becomes extremely ambiguous.
Are `facts'---in accordance with Bohr---characterized by a classical
description, or will the EP supply the definition of what can be regarded 
as a `fact'?
Or is a `fact' even regarded as independent of any theory like our data of
perception?
None of the alternatives fits in \cite{Haa 96}:
Was the concept of experience---in spite of the allegiance to 
Bohr---disregarded or dismissed as meaningless?
The EP has to take up the challenge of epistemology (which led Bohr astray to 
complementarism), especially the notion of
``indivisibility'' has to be justified anew (comparable to the meaning of space,
time, and causality in CP).
And the quest for realism can arise only from the demand to give this 
justification.
Instead, indivisibility is taken over from OQ for all events.
This offers a realistic picture indeed---for unobserved quantities;
but for observations---which was Bohr's main concern 
and touches profoundly the relation between nature and subject---the
EP offers nothing (just as well as CH). 

Contrary to this imperfection and ambiguity concerning epistemology,
indivisibility in the Whiteheadian approach (derived 
from an analysis of Zenon's paradoxes) is complemented by a subjectivistic 
metaphysics, which yields a theory of experience:
events (if causally tied) are also acts of experience.
As long as the relation between events and experience (connected with the
import of indivisibility) is not 
clarified---usually this is not in the responsibility of physicists---the
EP seems to be a naively realistic picture with an accidental similarity to a 
few of Whitehead's ``ideas''.
In order to introduce a realistic picture that ``transcend[s] Bohr's 
epistemology'' the efforts shouldn't be wasted on the status of unobserved 
quantities, even if positivistic criticism forces on such discussions.
Instead more relevant questions suggest themselves:
Is Bohr's concept of experience inevitable? 
What is hidden behind von Neumann's projection?
Was the interpretation that $\psi$ represents the 
(total) knowledge about a system provoked by this denial of the abstract
subject?
(An interpretation absolutely strange to Bohr.)

{\it 9. ``Does Quantum Physics force us to abandon the old picture of a real
outside world?''\/}
At least concerning nonrelativistic quantum physics there are more or less 
consistent theories without observer:
BM, CH, and SL (maybe a mixture of CH and SL will be added).
But the question for the role of the observer challenges the timeless view
human reason takes, which recognizes `nature' as the real outside world.
OQ, which is more and more made a scapegoat because 
of a misguided concept of nature, should be welcomed to develop a better 
understanding of CP.
Who except Bohr was ever truly inspired with this opportunity to get insights
in concealed prerequisites and with the quest to reconcile physical theories 
with cognitive faculties?
Proposals like historical or evolutionary pictures (with or without a
temporal logic) of quantum physics seem
to open up the possibility of giving experience of `nature' a temporary 
status and of `understanding' thinking as a process in `nature', 
but neither CH nor the EP get away from the timeless perspective offered by CP.
The challenge of physics and epistemology remains:
{\it how can thinking proceed in `time'?\/}
Reasonably, in \cite{Haa 96} no attempt is made to derive statements about
the mind from the EP.
But it will not take long until this challenge will rehash
the discussion about l'homme machine---now something like an information 
gathering and utilizing machine.

\end{document}